# Discharge dynamics in a cylindrical SDBD prototype reactor under ns-pulsed and sinusoidal AC operation

K. Giotis[1,2], D. Stefas[1], Y. Agha[1], H. Höft[3], X. Duten[1], P. Svarnas[2,a)], G. Lombardi[1], K. Gazeli[1,a)]

[1] *Laboratoire des Sciences des Procédés et des Matériaux (LSPM – CNRS), Université Sorbonne Paris Nord, Villetaneuse, UPR 3407, F-93430, France*

[2] *Plasma Technology Group, High Voltage Lab., Electrical & Computer Engineering Dept., University of Patras, 26504, Rion–Patras, Greece*

[3] *Leibniz Institute for Plasma Science and Technology (INP), Felix-Hausdorff-Straße 2, 17489, Greifswald, Germany*

[a)] *Authors to whom correspondence should be addressed: svarnas@ece.upatras.gr; kristaq.gazeli@lspm.cnrs.fr*

**Abstract**

We developed a prototype reactor generating surface dielectric barrier discharges (SDBDs) in ambient air, designed for consistent operation while preventing constructive material degradation. It features detachable stainless steel electrodes and quartz dielectric to ensure precise fabrication. The grounded electrode is fully immersed into transformer oil drastically suppressing undesired parasitic discharges. The device efficiently sustains ns-pulsed and AC discharges at 10 kHz, enabling fundamental studies of their electrical characteristics (applied voltage, induced current, electric power) and spatiotemporal dynamics (morphology, propagation length and velocity). The electric power (P) consumed exhibits a dissimilar non-linear increase with the rising peak voltage ($V_p$) in each case: P≈0.8–2.5 W for ns-pulsed ($V_p$=7–9 kV) and P≈0.9–5.3 W ($V_p$=7–10 kV) for AC operation. Using ICCD imaging, distinct ionization channels are recorded in the rising part of the pulsed voltage being detached from the driven electrode; during the voltage decrease, a glow-like discharge is formed remaining anchored on the driven electrode. The rising part of the AC voltage is characterized by erratic, elongated ionization channels in a filamentary form, the voltage drop featuring a glow-like behavior. During the rising and falling parts of the AC voltage, the discharge reaches maximum propagation lengths ($L_{max}$) of ≈12 mm and ≈7 mm, respectively, while remaining attached to the driven electrode. The corresponding maximum discharge velocities ($v_{max}$) are about $5\times10^2$ m/s and $3\times10^2$ m/s. For the ns-pulsed operation, $L_{max}$≈5 mm ($v_{max}$≈$5\times10^5$ m/s) and $L_{max}$≈3.5 mm ($v_{max}$≈$1.5\times10^5$ m/s) during the rising and falling parts of the voltage pulse, respectively. The SDBD dynamics generated with a ns-pulsed voltage is more reproducible than for the AC case allowing for the use of a 500 times smaller ICCD gate width (2 ns) and a more accurate description of the discharge's spatiotemporal development. This reactor is suitable for performing fundamental studies and understanding key SDBD features for various applications such as flow control, biomedicine and agriculture.

## I. INTRODUCTION

Surface dielectric barrier discharges (SDBDs) can yield non-equilibrium plasmas with increasing interest in fundamental studies, and promising applications in diverse fields. The latter include, but are not limited to, biomedicine [1], material processing [2], pollution control [3], *etc*. However, the main application points to the flow field control by means of plasma actuators. Comprehensive reviews have compiled the related state-of-the-art [4-7], where noteworthy examples refer to the control of the boundary layer separation [8], the laminar-to-turbulent transition delay [9], the stall improvement of an unmanned aerial vehicle [10], the performance of wind turbines [11], *etc*.

A typical SDBD reactor configuration comprises a dielectric layer in contact with two spatially separated electrodes, *i.e.*, a driven and a grounded one. In common designs, the grounded electrode is either encapsulated in the dielectric barrier itself or in another dielectric substrate, aiming to eliminate stray discharge formation in the vicinity of the grounded electrode. In terms of the shape of the dielectric barrier and the electrodes, orthogonal geometries are routinely adopted [4,5,7,12,13]. Despite the importance of such setups for numerous studies, their geometry noticeably differs from those used in many practical cases (see, *e.g.*, standardized airfoils [14-16]). Thus, more complicated geometries have been developed, producing plasma either of curved form on planar surfaces [17,18] or of such a form that osculates on curved surfaces where it propagates azimuthally (see [19-23] for cylindrical surfaces and azimuthal propagation).

Another key point in the design of SDBDs, directly affecting their features and efficiency, refers to the materials used. This becomes imperative due to the synergetic action of excited neutrals, radicals and charged species, UV photons, high electric and thermal fields, polarization effects, drift and displacement currents, *etc*. On top of that, in the case of SDBDs operating in ambient air, the materials are certainly subjected to humidity, ozone, ionized impurities, *etc*. In most conventional reactors, these factors result in material degradation and, in turn, in unstable plasmas, as it has been observed and studied extensively [24-34]. Hence both physical and chemical properties must be considered, like dielectric constant, dielectric strength, thermal capacitance, glass transition temperature, degree of oxidation, *etc*.

Concerning the dielectric barrier, in the case of the widely used polyamide films, it has been reported that they are readily etched/eroded [24-27], carbonized and swollen [28], *etc*., or in other words chemically and mechanically modified. This may lead to SDBD reactor failure within a few hours [25]. Similar degradation effects have been observed for other polymeric barriers too, *e.g.*, polymethyl methacrylate and polyvinyl chloride [29]. On the contrary, ceramic dielectrics (*e.g.*, quartz [25,26,30], aluminum oxide [25,26,31,32], borosilicate glass [27,29], *etc.*) appear much more resistant to such effects exhibiting long lifespan. Overall, any degradation of the barrier leads eventually to the alteration of the reactor specifications (*e.g.*, capacitance [24]) and plasma properties (*e.g.*, electric current [33]).

In addition, the role of the electrode material in the reactor and plasma stability is predominant. The usually employed copper (Cu) is susceptible to oxidation [25,27,30,31], while the same happens with aluminum [31,32] or alloys of noble metals (*e.g.*, AgPd and AgPt [25]) too. The use of pure noble metals (Ag [26,28], Au [34]) may tackle this issue, but they appear vulnerable to morphological changes. Other materials, like tungsten (W), have been proposed as a compromising solution [27]. Once more, like for the deterioration of the dielectric barrier, intense modification (chemical or mechanical) of the electrodes has an impact on the plasma features [25,26,30,31,34].

Furthermore, it is a common practice to encapsulate the grounded electrode in a dielectric medium, in an attempt to suppress any stray plasma formation. Otherwise, it may contribute to the degradation of the reactor [30], increase the power consumption [35], decrease the reactor efficiency [35], *etc*. Encapsulation processes refer either to attachment of polyamide films [25], acrylic plates [24], polyvinyl chloride plates [32], *etc*., or to resin curing (see epoxy resin [28], silicone resin [31], *etc.*). However, in both cases the residual air cavities or other imperfections are unavoidable, making the actual plasma volume rather ambiguous.

Once the SDBD reactor is designed, the driving voltage becomes a critical factor in the induced plasma features and in turn in its efficiency in eventual applications. The most common applied waveforms are AC and ns-pulsed driving waveforms [4]. Inherently, the SDBD dynamics depends on the driven voltage. This is directly reflected even in the general observation that an AC driving voltage leads to multiple, stochastic micro-discharges during the positive and negative slope of the waveform [4,5], whereas a ns-pulsed driving voltage leads in principle to two distinct, well repetitive discharges in the respective slopes of the high voltage pulse [4].

Accordingly, differently driven SDBDs are suitable for different applications. Thus, such discharges have been tested and compared for ozone production [36], ice mitigation [37], and flow field control [4] in both AC and ns-pulsed voltages. In ozone production, the ozone density notes higher values under the ns-pulsed driving voltage at the same consumed power [36]. In ice mitigation, the ns-pulsed operation found to be advantageous since it dissipates heat more efficiently [37]. Lastly, in the case of AC voltage, the flow field is governed by the momentum transfer from charged to neutral species, whereas in the case of ns-pulsed voltage, it is governed by fast gas heating [4]. Especially, each of these driving voltages finds a suitable application depending on the speed of the fluids that are involved [4].

The present work embraces all the above-discussed fundamentals to develop an SDBD reactor as a prototype for consistent studies of ns-pulsed and AC plasma dynamics in atmospheric air. Thus, on the one hand, the reactor: (i) adopts a cylindrical geometry; (ii) is based on a dielectric barrier made of quartz; (iii) uses electrodes consisting of stainless steel; and (iv) has a grounded electrode immersed in high dielectric strength transformer oil, drastically eliminating stray discharges compared to other methods. In addition, the reactor comprises fully detachable distinct elements, leaving no space for clumsy handling during its fabrication (contrary to the cases where adhesive tapes, resins, terminations, *etc.* are employed). On the other hand, the reactor is driven by either ns-pulsed or AC sinusoidal voltage (both in the kHz range), and transient electrical and optical effects are analyzed. Hence, comparison between induced currents and consumed power, discharge patterns formed within various time windows, propagation length and velocity are achieved, for both operational modes.

## II. EXPERIMENTAL SETUP AND METHODS

**Figure 1** provides a detailed technical drawing and a photograph of the reactor. The dielectric barrier is a cylindrical tube (2 mm wall thickness) made of quartz (TQS®). Both electrodes are made of stainless steel (Acciaierie Valbruna S.p.a.; grade 304). The driven one (outer) forms a hollow truncated cone of 10 mm in height. The upper/lower edges of this electrode have a thickness of 1/0.1 mm. The grounded electrode has the form of a cylindrical tube (1 mm wall thickness), it is housed in the quartz in such a way that its surface osculates the inner surface of the quartz, along its entire length, and it is fully immersed in transformer oil (Cargill®; Envirotemp™ 200 Fluid). In this study, there exists no vertical gap between the driven and the grounded electrode, *i.e.*, the lower edge of the former is at the same level as the beginning of the latter. O-rings (Viton®) are used for oil confinement in the quartz tube while the top cap of the tube (Teflon™) brings a heat/pressure release opening.

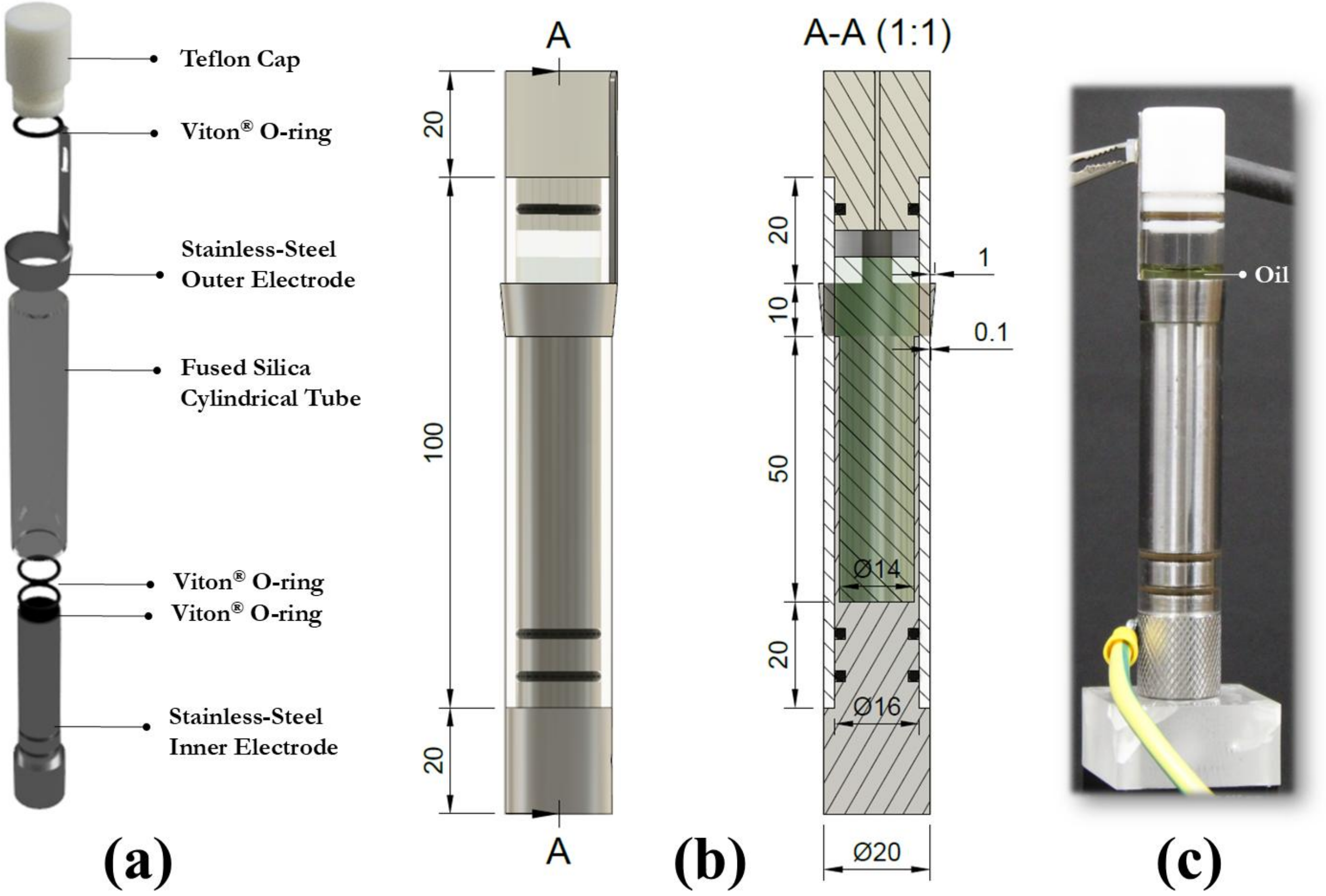

**Figure 1.** *SDBD reactor:* ***(a)*** *Disassembled components.* ***(b)*** *Technical drawing.* ***(c)*** *Photograph. Insulating oil is distinguishable in (b) and (c).*

**Figure 2** presents the experimental setup used to investigate the SDBD. The reactor is powered by two independent power supplies. The first generates square pulses and is based on a high voltage DC unit (HCN700-20000; F.u.G. Elektronik GmbH), a solid-state high voltage chopper (PVX-4110; Directed Energy Inc.), and a signal generator (DG645; Stanford Research Systems). Voltage amplitude up to 10 kV (peak), pulse frequency up to 10 kHz, pulse rising/falling time of about 60 ns, and pulse width (FHWM) down to 150 ns can be produced (plateau 110 ns). The second power supply is a fully standalone sinusoidal one (PlasmaHTec®), yielding voltage amplitude up to 15 kV (peak), real power up to 500 W (mean), and constant frequency of 10 kHz with a total harmonic distortion down to 1%. Herein, this frequency is set to both power supplies. The waveforms of the driving voltage and the total current (drift and displacement) are monitored on a digital oscilloscope (Teledyne LeCroy; WaveSurfer10; 1 GHz; 10 GS $s^{-1}$) by means of high voltage passive probe (Tektronix; P6015; DC – 75 MHz) and wideband current transformer (Magnelab; CT-D2.5-BNC; 1.2 kHz – 500 MHz), respectively. Any signal propagation delay, due to both probes, has been carefully compensated (accuracy <1 ns).

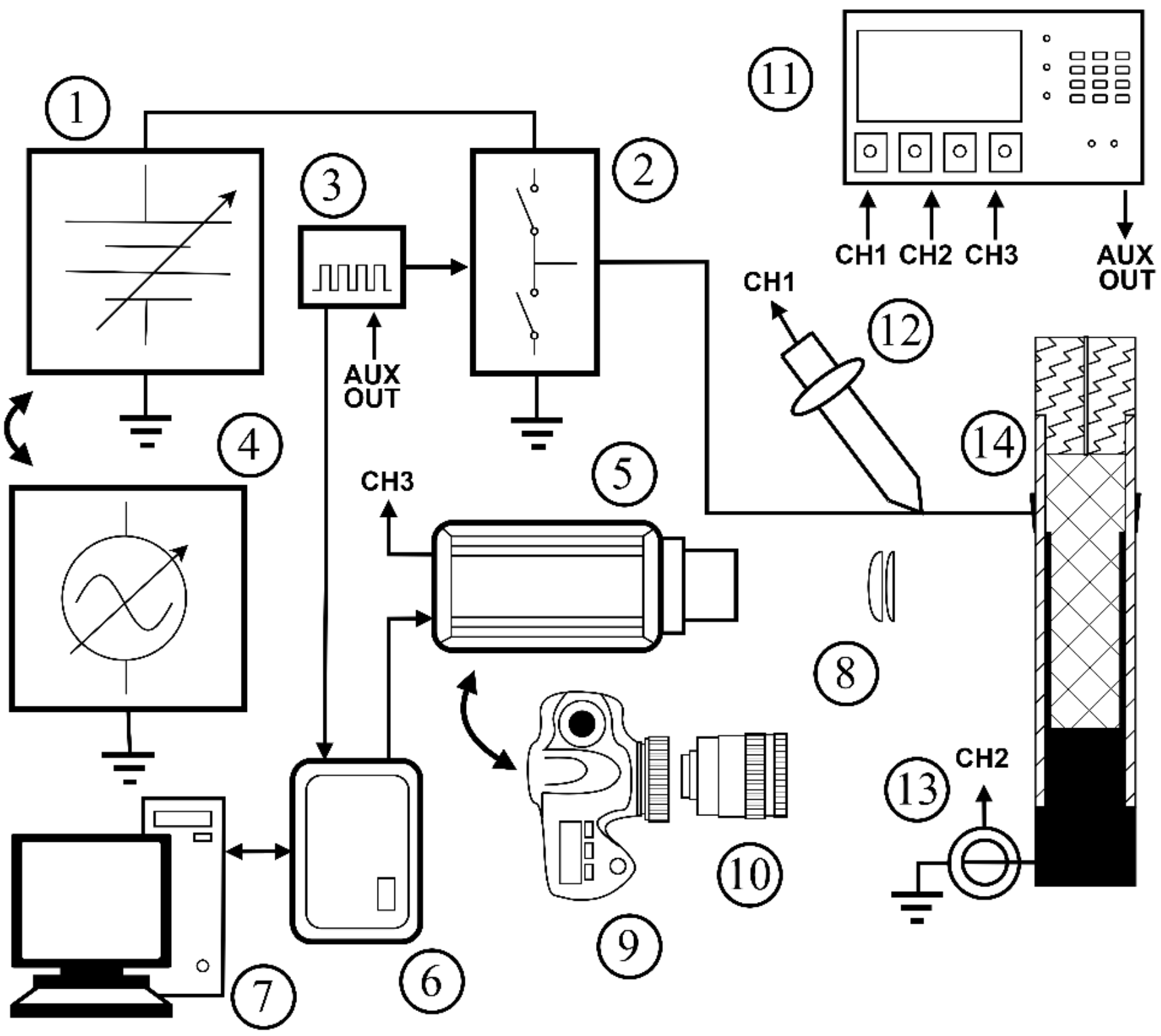

**Figure 2.** *Conceptual diagram of the experimental setup: (1) high voltage DC power supply; (2) high voltage chopper; (3) signal generator; (4) high voltage AC power supply; (5) ICCD camera; (6) ICCD controller; (7) personal computer; (8) UV lenses; (9) DSLR camera; (10) objective lens; (11) oscilloscope; (12) high voltage probe; (13) current transformer; (14) SDBD reactor section. Interchangeable parts: (4) ↔ (1)-(2) and (9)-(10) ↔ (5)-(8).*

The recorded voltage and current waveforms allow for the calculation of the consumed electric power, *i.e.*:

$$\bar{P}_{DBD}^{Pulsed} = \frac{1}{T}\int_{\tau}^{\tau+\Delta t} \boldsymbol{v}(t) \times \boldsymbol{i}(t)\mathrm{d}t \qquad (1)$$

$$\bar{P}_{DBD}^{AC} = \frac{1}{10}\sum_{n=1}^{10}\left[\frac{1}{T}\int_{T} v(t) \times i(t)\mathrm{d}t\right]_{n} \qquad (2)$$

where $T$ stands for the driving voltage period, $\Delta t$ for a time interval adequate to include the entire high voltage square pulse, the bold text quantities refer to waveforms recorded in the average acquisition mode of the oscilloscope (100 waveforms), the regular text quantities imply normal (not averaged) acquisition mode, and the quantity $n$ mentions the number of the independent records used. Two experimental sets have been conducted on different dates in controlled laboratory ambience (20°C temperature; 40–50% relative humidity).

The discharge dynamics is studied by means of an ICCD camera system (PIMAX-1K-RB-FG-P43; Princeton Instruments). The system includes a fast gated intensifier (2 ns minimum gate width), an image sensor (Marconi CCD47-10; 1024×1024 px$^2$; 13 μm pixel size), and a controller (ST-133; PTG as internal pulser). Two ultraviolet fused-silica, plano-convex lenses (LA4184; 500 mm, LA4579; 300 mm; Thorlabs) are mounted back-to-back (detail 8 in **Fig. 2**) to reduce lens aberration as much as possible, providing an effective focal length $f_{\mathrm{eff}}$ of 187.5 mm. Using this combination, a $2f_{\mathrm{eff}}$–$2f_{\mathrm{eff}}$ configuration is then formed for side-on observations of the SDBD. In this way, wavelength-integrated images (dominant emissions are attributed to excited $N_2$ and $N_2^+$ [38]) of about 1:1 scale are projected to the image sensor. The probed area corresponds to a square of 13.8 × 13.8 mm$^2$. Two sets of experiments are performed. The captured images are visualized on a symbolic logarithmic scale. Physical quantities (*i.e.*, propagation discharge length and velocity) are finally extracted by employing systematic digital processing techniques.

The synchronization of the ICCD with the driving voltage waveform differs between the ns-pulsed and the AC case. In the former case, the delay generator triggers both the power supply and the ICCD. Various delays are used for acquiring records over the entire period of the high voltage pulses. The resultant jitter between the high voltage pulse and the camera gate is measured to be less than 1 ns. In the latter case, however, the delay generator is externally triggered by the auxiliary output of the oscilloscope. This output reflects the oscilloscope triggering event with respect to the AC waveform. The subsequent signal of the generator triggers the ICCD. Once again, various delays are used for acquiring records over the entire period of the AC high voltage. In the AC case, the resultant jitter between the high voltage and the camera gate is measured to be around 100 ns. Finally, conventional photos of the visual appearance of the plasma are recorded using a digital lens reflex (DSLR) camera (CANON EOS 4000D; 55 mm focal length; f/5.6 f-stop; 1/30 s integration time; ISO-6400).

A dedicated numerical routine is developed for the analysis of the ICCD images (Python-based code) so that the propagation length and velocity of the discharge can be evaluated by means of a settled algorithm, as follows; see [30] and [39] for similar reported processes. The raw ICCD images (see typical in **Figure 3a**) contain background noise which is here eliminated by a thresholding procedure (**Fig. 3b**). Subsequently, a 3×3 median filter is applied to the images for further denoising (**Fig. 3c**). Then, since the advancing contour of the visualized discharge pattern has a curved, not defined geometry, a per row integration of the light intensity (pixel values) is employed to achieve the emission intensity profile along the discharge propagation axis (vertical). Two such profiles are depicted in the right inset of **Fig. 3d**, which correspond to successive propagation moments, and they are further compared in **Fig. 3e**. The displacement (propagation) position of the discharge front is here identified as the furthermost vertical pixel where the profile curve extinguishes to 5% of its peak value (see star symbols in **Fig. 3e**). In case of absence of a pixel assigning an intensity at this exact percentage, a linear interpolation is applied among adjacent pixels. Afterwards, the position of the abovementioned 5%-pixel is converted to a physical position due to the relation of 0.0135 mm $px^{-1}$ being found by conventional calibration process. Those critical 5%-pixels of the propagation position from successively captured images are eventually considered for the determination of the discharge propagation length and velocity.

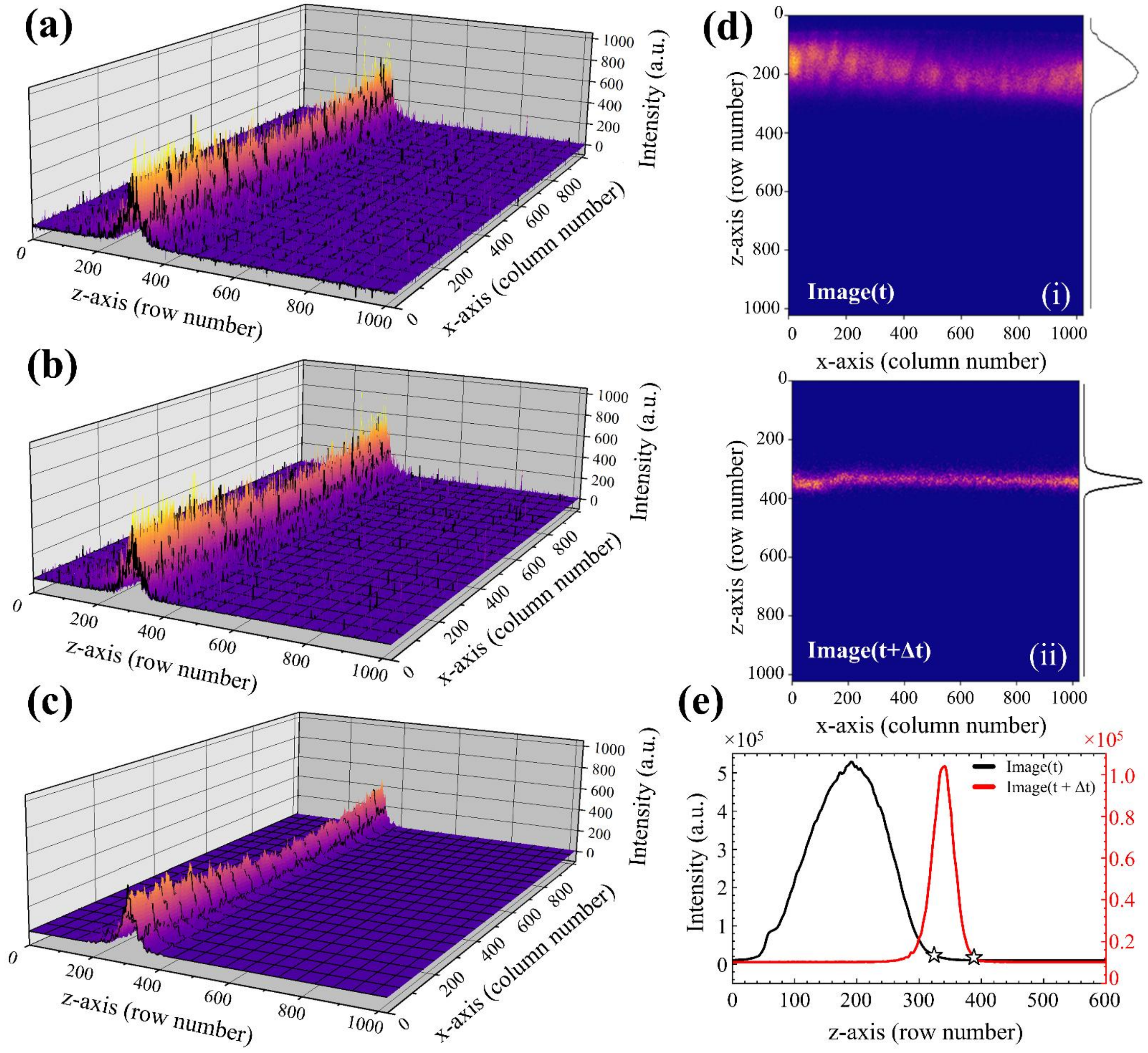

**Figure 3.** ***(a)*** *Raw ICCD image.* ***(b)*** *Threshold application (pixel intensity ≤ 10).* ***(c)*** *3×3 median filter application.* ***(d)*** *Resultant images from two different time instants and rows-integrated intensity profiles.* ***(e)*** *Intensity profiles seen in (d) where the 5%-pixels are marked by the star symbols.*

## III. RESULTS AND DISCUSSION

**Figure 4** presents the driving voltage waveforms employed here, as well as the corresponding induced total current signals. It is noted that, in both voltage cases the total currents result from the displacement (due to the polarization of the quartz) and the proper discharge-current components (due to the drift of the plasma charged species). The decomposition of the total current to these components remains a challenge in the case of SDBDs since the capacitance of the reactor varies during the discharge development [40]. However, recorded current waveforms may be related to the discharge dynamics [4,5,30,40-45] and the results given below support this relation.

The ns-pulsed voltage shown in **Fig. 4a** has a rising/falling time of about 35 ns (10%-90%; experimentally measured during SDBD operation) and a plateau lasting about 110 ns, while its frequency is maintained at 10 kHz. Due to the steep rising/falling slopes of the ns-pulsed voltage ($^{\Delta V}/_{\Delta t}$≈200 V/ns), a strong capacitive current is generated. Besides, the total induced current exhibits two distinct and reproducible peaks, one during the rising (magnitude up to 2 A) and a second one during the falling edge of the voltage pulse. The negative current peak appears always weaker than its positive counterpart. Although the extraction of the actual discharge current is challenging, the instants of discharge inception and extinction within the timespan of the total current (see dashed lines in **Fig. 4a**) can be identified from ICCD images (studied below). Thus, one can estimate the duration of the discharge current induced within the rising and falling parts of the ns-pulsed voltage, here being ≈30 ns and ≈70 ns, respectively. These findings are corroborated by reference [43], where a linear SDBD arrangement with a grounded electrode encapsulated in epoxy resin was studied.

On the other hand, the AC voltage (**Fig. 4b;** rising/falling parts: $^{\Delta V}/_{\Delta t}$≈0.5 V/ns) is a sinusoidal one (total harmonic distortion <1%) at the same as above frequency, *i.e.*, 10 kHz. In this case, the discharge current is clearly distinguishable from multiple mA-peak impulses superimposed to a weak capacitive current. For instance, more than 150 impulses are recorded in the rising part of the AC waveform. Their peak values can reach up to 70 mA (see inset in **Fig. 4b**; noise threshold considered to be around 6 mA), being up to ≈10 times larger than during the falling part. From all these impulses, an average peak intensity of ≈18 mA and a total duration of around 20 ns are measured. These values are close to those reported in reference [35], referring to a planar SDBD actuator operated at voltage amplitudes $V_p$≥12 kV for which the grounded electrode was encapsulated in epoxy resin. The different total currents recorded between ns-pulsed and AC excitations are due to the dissimilar type and absolute value of $^{\Delta V}/_{\Delta t}$ of the two voltage waveforms.

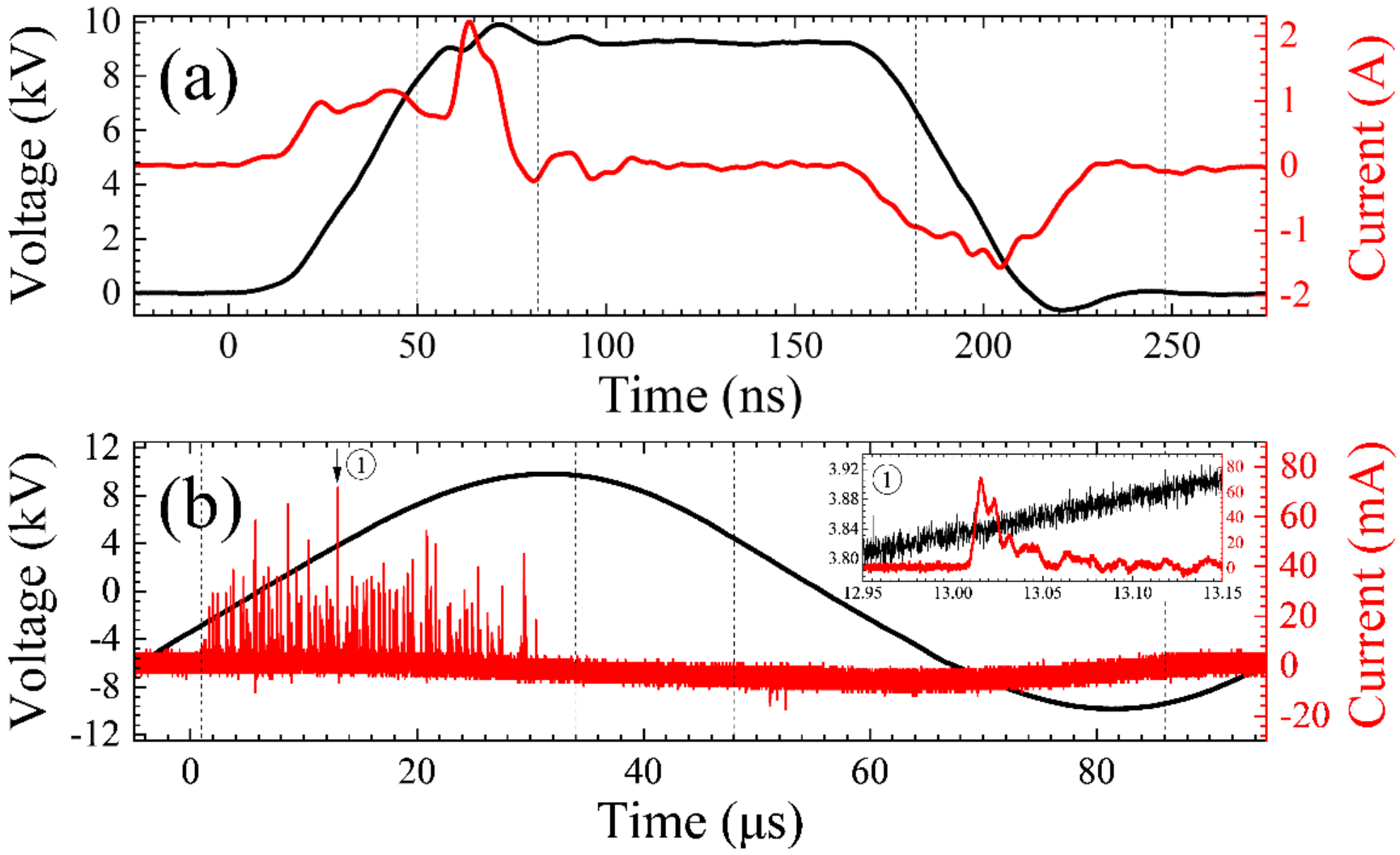

**Figure 4.** *Driving voltage and induced total current waveforms **(a)** ns-pulsed (only a zoom on the pulse is shown since the applied voltage and induced current are zeroed during the rest of the HV period) and **(b)** continuous AC operation (the inset displays a zoom around the largest current impulse captured; the timescale displayed is similar to that of the voltage pulse shown in (a)). The dashed lines in (a) and (b) indicate the instants of the discharge inception and extinction during the rising and falling parts of the corresponding voltage waveforms. These instants are identified using time-resolved ICCD images (see **Figures 7-10**).*

**Figure 5** displays the mean electric power consumed as a function of the amplitude ($V_p$) of the driving voltage in both cases. This is a convenient representation especially for flow control applications, since the induced body forces depend on the applied voltage in a similar manner to the consumed power [7,40,46]. Overall, average power values between 0.5 and 5.3 W are measured for both voltages ($V_p$ range studied: 7–10 kV). In both cases, a higher amplitude results in increased power consumption, which is a rational tendency [30,40,43,44,46-48]. Using the powers measured in **Fig. 5** and the circumference of the cylindrical electrode (≈63 mm), the dissipated power per unit length (i.e., W/cm) of the powered electrode of the SDBD can be calculated, giving 0.14–0.4 W/cm for the ns-pulsed case ($V_p$=7–9 kV) and 0.14–0.85 W/cm for the AC case ($V_p$=7–10 kV), being close to corresponding values measured in planar SDBDs [47,49,50]. The narrower error bars in **Fig. 5a**, as compared to those in **Fig. 5b**, are probably due to the higher reproducibility of the current pulses in the ns-pulsed case (see **Fig. 4** and discussion there). Although the discharges produced by the two different voltage waveforms are not directly comparable, the higher power consumed under AC operation, within the present operational windows, is notable.

Furthermore, a preliminary rough approximation of the dependence of the consumed power on the voltage amplitude in each case can be done by fitting the experimental results with the red curves shown in **Fig. 5**. This leads to a power relation proportional to the term $V_p^a$, where $a \approx 3.8 \pm 0.3$ and $5.4 \pm 0.2$ for the ns-pulsed and the AC operation, respectively. Similar relations have been proposed in the literature for planar SDBDs driven only by AC high voltages, finding however $a \approx 2 - 3.5$ [40,46-48], *i.e.*, a lower value than in the present setup. The fact that values of $a$ are larger than 2 is because, besides the power losses in the dielectric material (being proportional to $V_p^2$), additional power is consumed to form the various discharge channels [46,47]. Numerous reported SDBDs usually refer to planar configurations typically operating below 10 kHz and utilizing common dielectrics (such as Kapton®, plexiglass, Teflon and glass) and electrodes (e.g., copper foil tapes). These materials can be prone to aging, e.g., due to the presence of air cavities between the grounded electrodes and the dielectrics, leading in undesired partial discharge formation. For instance, fast oxidation of copper stripes was reported in a planar SDBD (using epoxy resin to encapsulate the grounded electrode), causing a noticeable decrease in the total consumed power over time [30]. The present cylindrical SDBD is operated at 10 kHz and encompasses a non-conventional but highly-efficient encapsulation method of the grounded electrode (i.e., using transformed oil with high dielectric strength). Besides, this SDBD uses fused silica as dielectric which has demonstrated a noticeable stability and sturdiness in a planar SDBD driven by AC voltage [30]. This configuration drastically suppresses air cavity formation and eliminates parasitic discharges, thus allowing consistent operation over long time and more efficient power transfer to the discharge. This fact and the relatively high operating frequency used here could partly explain the higher value of $a$ found in **Fig. 5b** when compared to planar SDBDs [40,46-48]. Another reason for this difference could be due to the fact that the AC voltage range studied in **Fig. 5b** is smaller compared to those in references [40,46-48]. Operating the SDBD in a wider range (not feasible here), e.g., 7 kV $\leq V_p \leq$ 20 kV, would provide a larger set of experimental points to extract a more representative power law. This procedure could possibly lead to a value of $a$ within (or close to) the range $a \approx 2 - 3.5$ reported in [40,46-48]. Other factors that may restrict the generalization of these values may be attributed to the differences in the power supplies used, SDBD structural features, and electrical conditions applied (voltage type and amplitude, frequency, etc.) [40]. An exhaustive study of the effect of the voltage amplitude/frequency on the total consumed power of the cylindrical SDBD is left for future work.

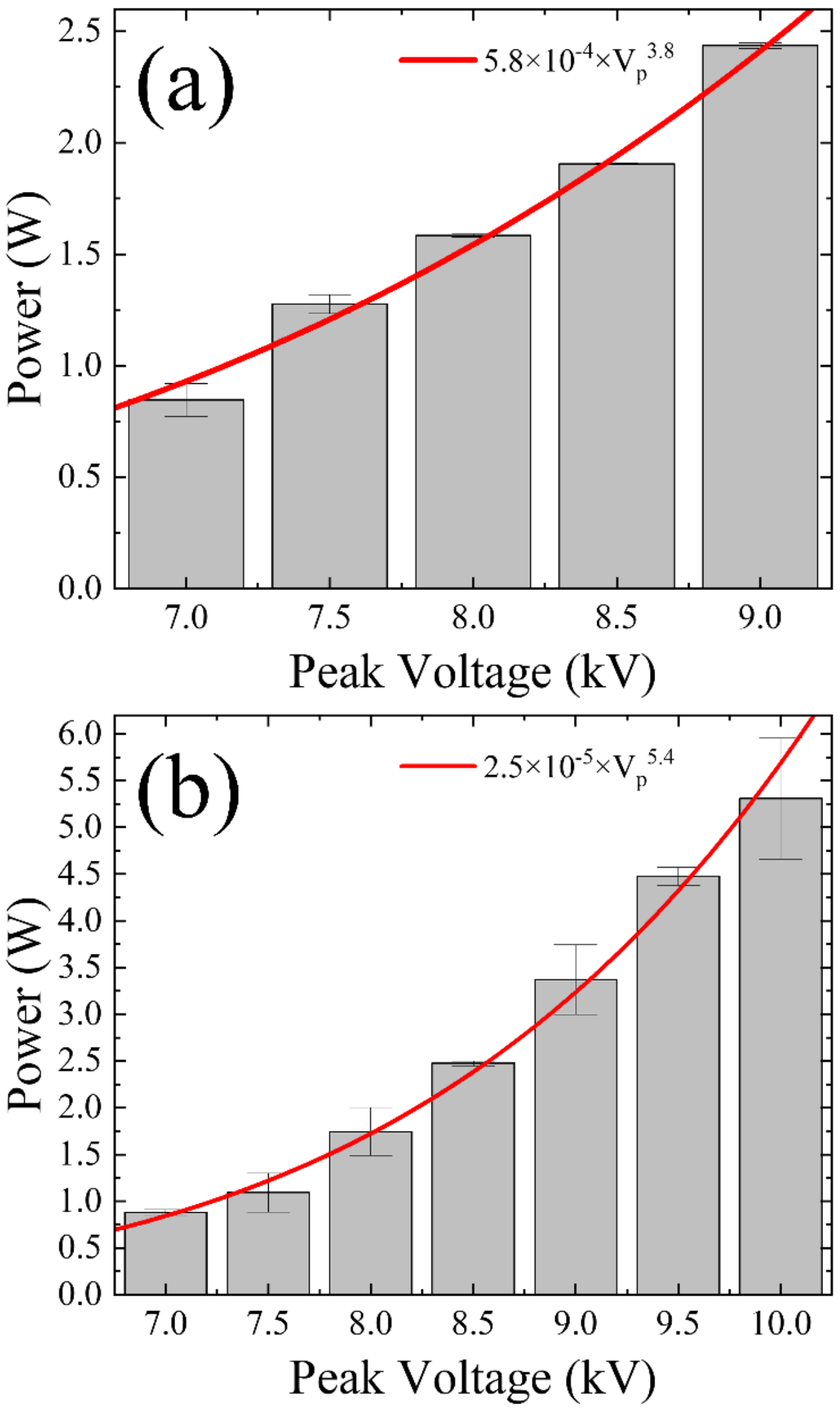

**Figure 5.** *Mean electric power (P; grey columns) as a function of the voltage amplitude ($V_p$):* ***(a)*** *ns-pulsed and* ***(b)*** *continuous AC operation. The experimental data are best fitted with the red lines described by the following formulae:* $P = 5.8 \times 10^{-4} \times V_p^{3.8}$ *(ns-pulsed operation) and* $P = 2.5 \times 10^{-5} \times V_p^{5.4}$ *(AC operation).*

**Figure 6** depicts the visible time-integrated patterns of the discharges, as they are produced by the two different voltage waveforms. In a qualitative manner, in the ns-pulsed case this pattern consists of well separated emissive, quite wide, channels starting from narrow zones localized on the driven electrode and propagating onto the dielectric surface downstream of the electrode. In the AC voltage, the corresponding pattern may be separated into two parts: a first one consisting of numerous, intense, “corona-like”, filamentary discharges spread around the circumference of the driven electrode, and a second part where these numerous discharges are diffused to form a “glow-like” circular zone, propagating onto the dielectric surface downstream of the electrode. The features of these patterns are related to the change of the polarity of the powered electrode from positive to negative and are below validated by more rigorous considerations.

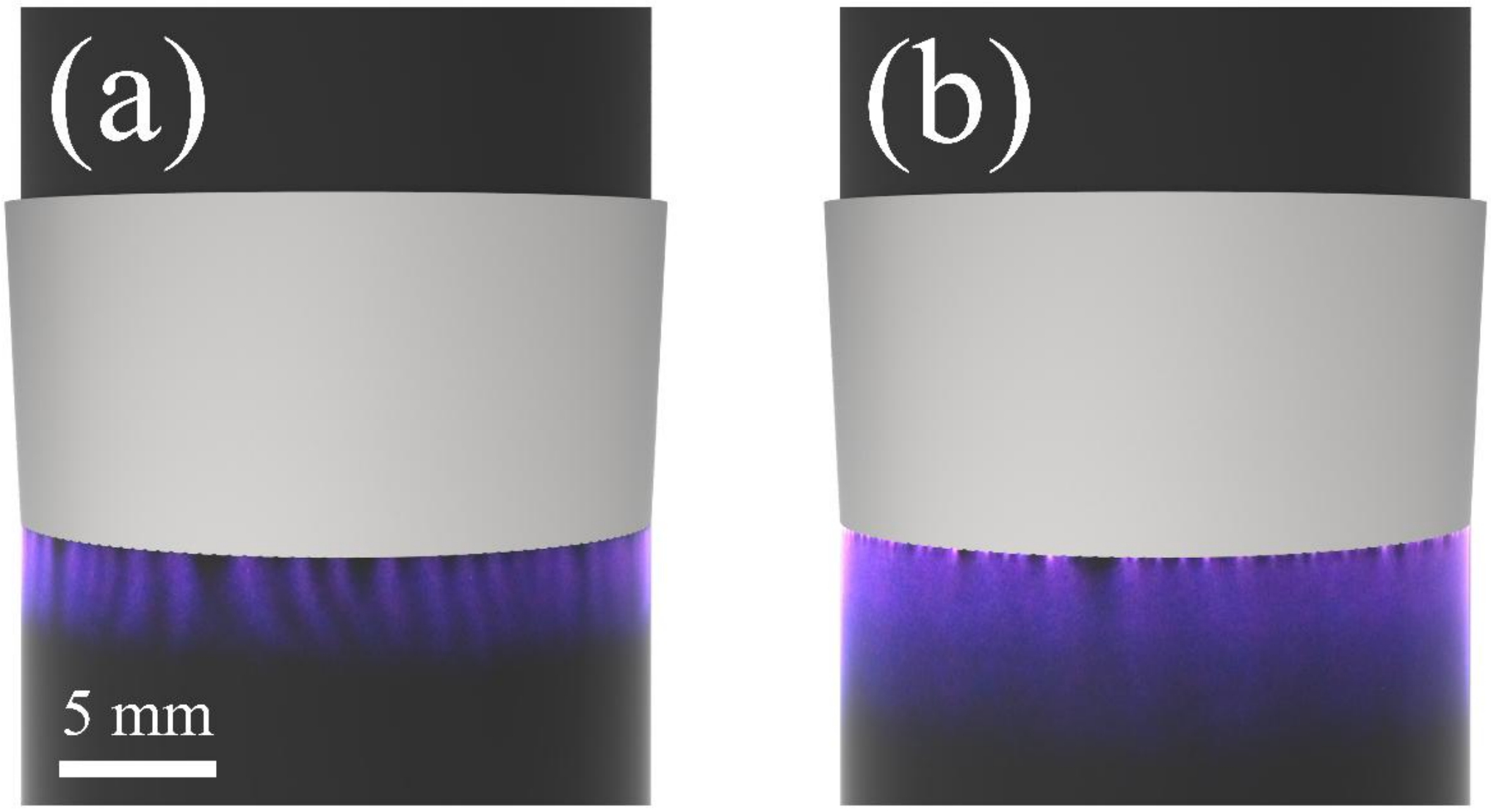

**Figure 6.** *SDBD illustration based on conventional photos of the discharge (exposure time: 1/30 s) and rendered parts of the reactor close to the electrodes:* ***(a)*** *ns-pulsed operation (9 $kV_p$ ) and* ***(b)*** *continuous AC operation (10 $kV_p$ ).*

**Figure 7** displays ICCD images captured within the rising and falling parts for both the ns-pulsed (**Fig. 7a**) and the AC operation (**Fig. 7b**). These offer better insights into the corresponding discharge patterns formed (see **Fig. 4** for time correlation with voltage waveforms). Hence, one single ICCD image (1 accumulation) is captured in such a way that integrates either the rising (**Figs. 7ai** and **7bi**) or the falling (**Figs. 7aii** and **7bii**) part of each waveform. Overall, **Fig. 7** validates the patterns qualitatively discussed on **Fig. 6** and can be compared to previously presented fast imaging results from SDBDs driven by ns-pulsed and AC voltages. For instance, Benard *et al.* [49] studied a planar SDBD driven by a similar ns-pulsed power supply as in our case. The discharge pattern exhibited a clear filamentary structure during the rising and falling voltage parts. In fact, during the rising voltage part, numerous streamers were generated over the length of the powered electrode, being superimposed to a weaker corona-like discharge background and exhibiting propagation lengths up to 4 mm. These in general agree with our results shown in **Fig. 7ai**. During the falling voltage part, the discharge development is due to the release of previously-accumulated positive charges on the dielectric surface. This also exhibited intense filaments, presenting however a curly structure possibly due to a non-uniform charge deposition on the dielectric. Compared to reference [49], the discharge generated during the falling voltage part in our case (**Fig. 7aii**) does not contain clearly-separated intense filaments and appears to be more uniformly-spread along the dielectric surface, as in reference [41]. This difference may be attributed to the lower voltage amplitude/width/rise time used here (9 kV/150 ns/35 ns instead of 10 kV/200 ns/50 ns in [49]), and the structural differences between the two SDBD reactors. In the AC case, the results obtained corroborate those presented in [50]. The ICCD images recorded support the generation of positive streamers during the rising voltage part (**Fig. 7bi**), extending stochastically along the dielectric surface and reaching lengths up to about 12 mm. In the falling voltage part (**Fig. 7bii**), however, a glow-like discharge is generated resulting in a more uniform pattern over the dielectric surface, reaching lengths up to 8 mm.

Here, it is underlined that the intensities of the ICCD images shown in **Fig. 7** are not directly comparable between them. However, using space-time-resolved ICCD imaging key discharge quantities such as the ionization wave velocity can be extracted, as presented and discussed below.

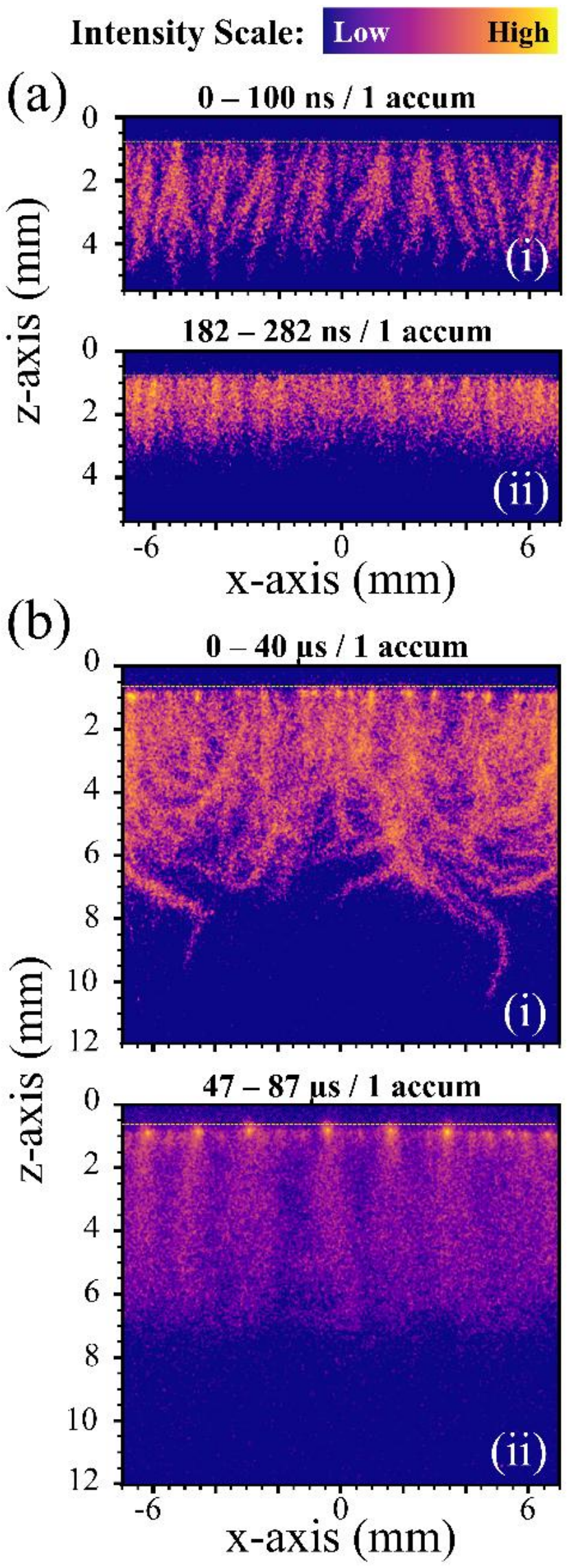

**Figure 7.** *ICCD fast imaging of the discharge; see* ***Fig. 4*** *for time correlation with voltage waveforms).* ***(a)*** *ns-pulsed operation (9 kV$_p$):* ***(i)*** *rising part (100 ns gate width, 255 gain) and* ***(ii)*** *falling part (100 ns gate width, 255 gain).* ***(b)*** *AC operation (10 kV$_p$):* ***(i)*** *rising part (40 µs gate, 255 gain);* ***(iv)*** *falling part (40 µs gate, 255 gain). The gate activation within the voltage period is defined by the times given on the top of each frame. The yellow dashed line denotes the boundary of the electrode gap (section II). Please note that the x-axis shown in the images refers to the length of the cylindrical chord (13.8 mm) seen by the camera, and not to the length of the corresponding arc (15.2 mm).*

**Figure 8** refers to ns-pulsed voltage only and considers the corresponding discharge dynamics, distinguishing the rising and falling parts (see **Fig. 4** for time correlation with voltage waveforms). For that, time resolved images with the best possible temporal resolutions of the ICCD are captured either with or without frame accumulation during both voltage parts. Accordingly, **Fig. 8a** (gate 2 ns) provides images accumulated over 100 periods during the voltage rising (**ai – aiii**) and falling (**aiv – avi**) parts. On the other hand, **Fig. 8b** (gate 5 ns) provides single shot, not accumulated, images during the voltage rising (**bi – biii**) and falling (**biv – bvi**) parts. Independently of the accumulation number, the results unveil distinct characteristics of the discharge propagation as the externally applied electric field increases or decreases. As per the results of **Fig. 7**, during the voltage increase, distinct ionized channels (resembling cathode-directed streamers) are developed and propagate downstream of the driven electrode. However, here it is additionally revealed that the discharge developed during the rising voltage part is detached from the driven electrode, which is not evident in **Fig. 7a**. Then, during the voltage decrease, a glow-like discharge (resembling cathode-sheath) is formed, propagating downstream of the driven electrode and remaining anchored on that. These observations are in good agreement with previously reported results in planar SDBDs [30, 41-43].

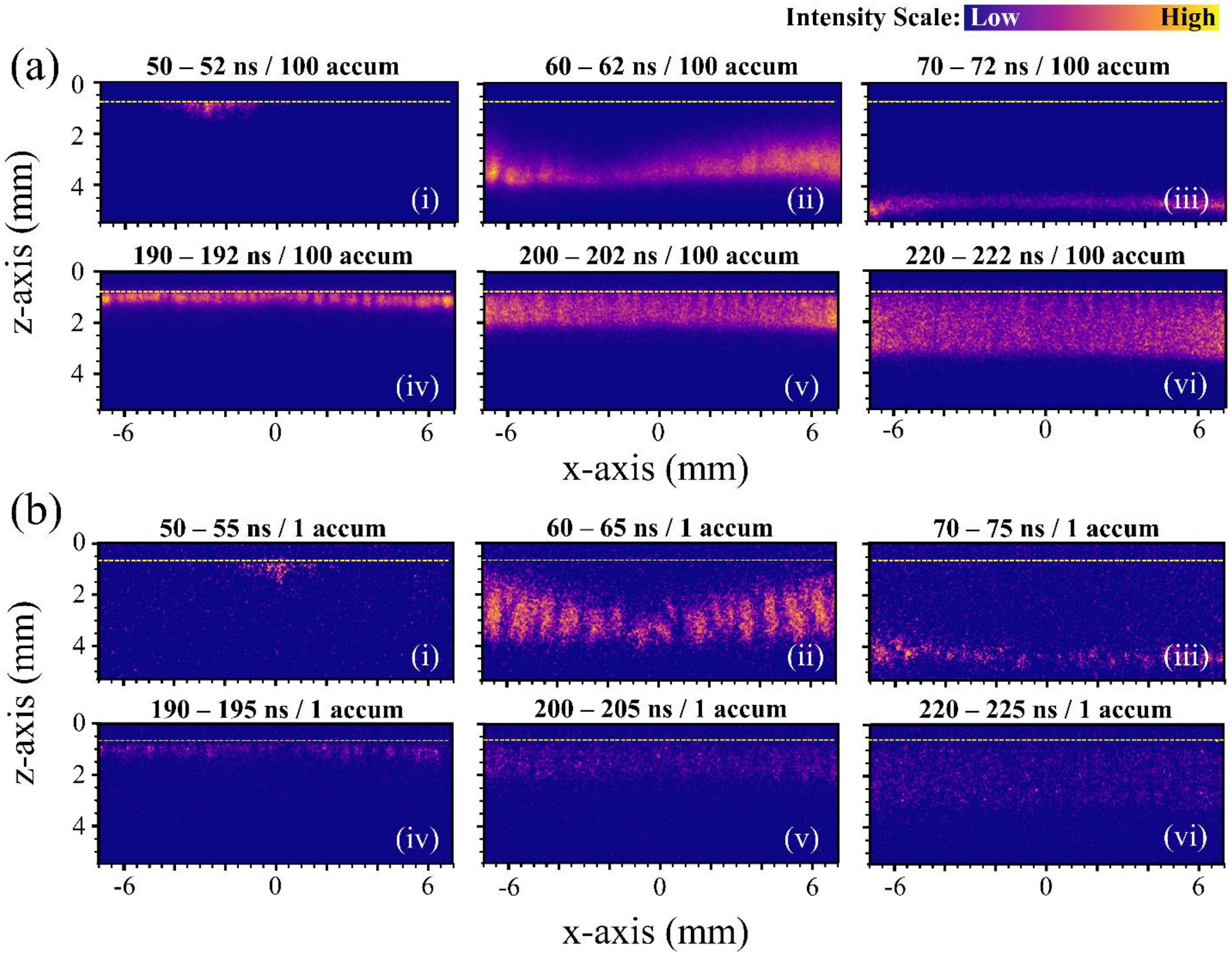

***Figure 8.*** *ICCD fast imaging of the discharge in ns-pulsed operation (9 $kV_p$); see Fig. 4 for time correlation with voltage waveforms.* ***(a)*** *2 ns gate, 100 accumulations and 255 gain:* ***(i), (ii), (iii)*** *rising part;* ***(iv), (v), (vi)*** *falling part.* ***(b)*** *5 ns gate, 1 accumulation and 255 gain:* ***(i), (ii), (iii)*** *rising part;* ***(iv), (v), (vi)*** *falling part. The gate activation within the voltage period is defined by the times given on the top of each frame. The yellow dashed line denotes the boundary of the electrode gap (section II).*

**Figure 9** shows the corresponding to **Fig. 8** results in the case of AC operation (see **Fig. 4** for time correlation with voltage waveforms). In other words, time resolved images are captured either with or without frame accumulation during both voltage parts. Accordingly, **Fig. 9a** (gate 1 μs) provides images accumulated over 100 periods during the voltage rising (**ai – aiii**) and falling (**aiv – avi**) parts. On the other hand, **Fig. 9b** (gate 1 μs) provides single shot, not accumulated, images during the voltage rising (**bi – biii**) and falling (**biv – bvi**) parts. Those latter images demonstrate that the pattern shown in **Fig. 7bi** (voltage rising part) is due to the development of erratic, elongated ionization channels in a filamentary form, and that shown in **Fig. 7bii** (voltage drop) is due to weak ionization, both being in a close analogy to the numerous impulses observed on the current waveform (**Fig. 4b**). In addition, as the voltage approaches the peaks of the sinusoid (positive or negative) ionization extends downstream of the driven electrode (**Figs. 9a** and **9b**). Apart from a recent report devoted to the investigation of an AC-driven planar SDBD using time-resolved ICCD imaging [51], the present observations comprise a detailed demonstration of the quite stochastic spatiotemporal fingerprints of SDBDs operating with AC excitation. This point is further emphasized when the discharge dynamics under AC excitation is compared to that produced using a ns-pulsed voltage.

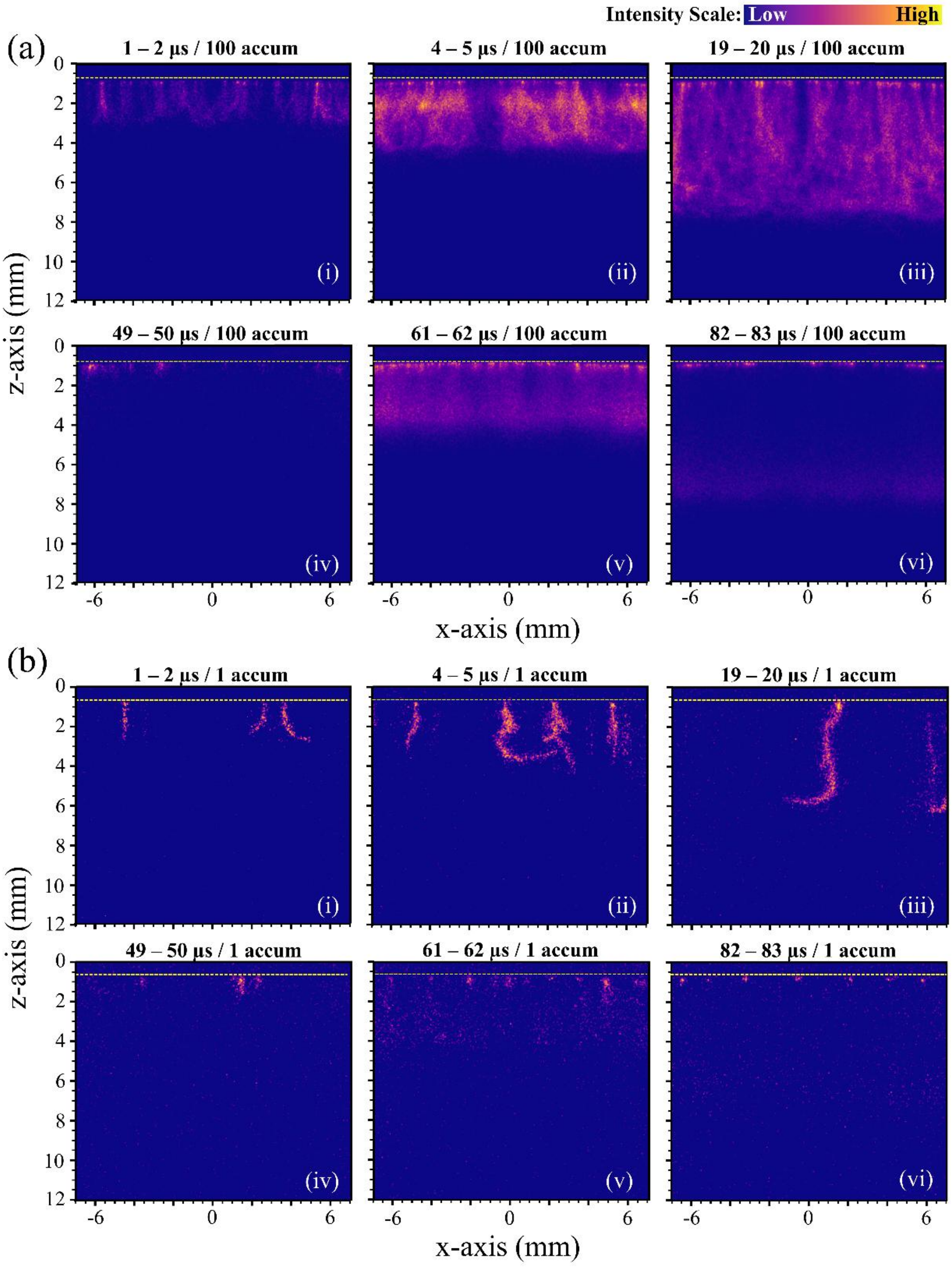

***Figure. 9.*** *ICCD discharge imaging in AC operation (10 $kV_p$); see Fig. 4 for time correlation with voltage waveforms.* ***(a)*** *1 μs gate, 100 accumulations and 255 gain:* ***(i), (ii), (iii)*** *rising part;* ***(iv), (v), (vi)*** *falling part.* ***(b)*** *1 μs gate, 1 accumulation and 255 gain:* ***(i), (ii), (iii)*** *rising part;* ***(iv), (v), (vi)*** *falling part. The gate activation within the voltage period is defined by the times given on the top of each frame. The yellow dashed line denotes the boundary of the electrode gap (section II).*

The discharge dynamics during propagation along the dielectric surface of the present reactor is better demonstrated by **Figs. 10** and **11**, which refer to the ns-pulsed and AC operation respectively (see **Fig. 4** for time correlation with voltage waveforms). These figures show row-integrated UV-NIR intensity profiles (see **Fig. 3** and analysis there), supporting many of the discussions in the previous paragraphs and unveiling new features. The intensity graphs are normalized to the maximal intensity recorded in each case (ns-pulsed and AC).

Concerning the ns-pulsed discharge and the rising part of the voltage **(Fig. 10a)**, an ionizing front propagates along the *z*-axis downstream, *i.e.*, downstream of the driven electrode (denoted by the light gray line in the figure). This front is intensified at 52 ns in **Fig. 10a** and extinguishes progressively within the next 30 ns (approx.). During this propagation the ionization channel is clearly detached from the driven electrode. As regards the falling part of the voltage pulses (**Fig. 10b**), a prolonged discharge channel seems to be built up, without a clear detachment from the driven electrode. This channel is formed, intensified and, eventually, extinguished within a period of about 70 ns.

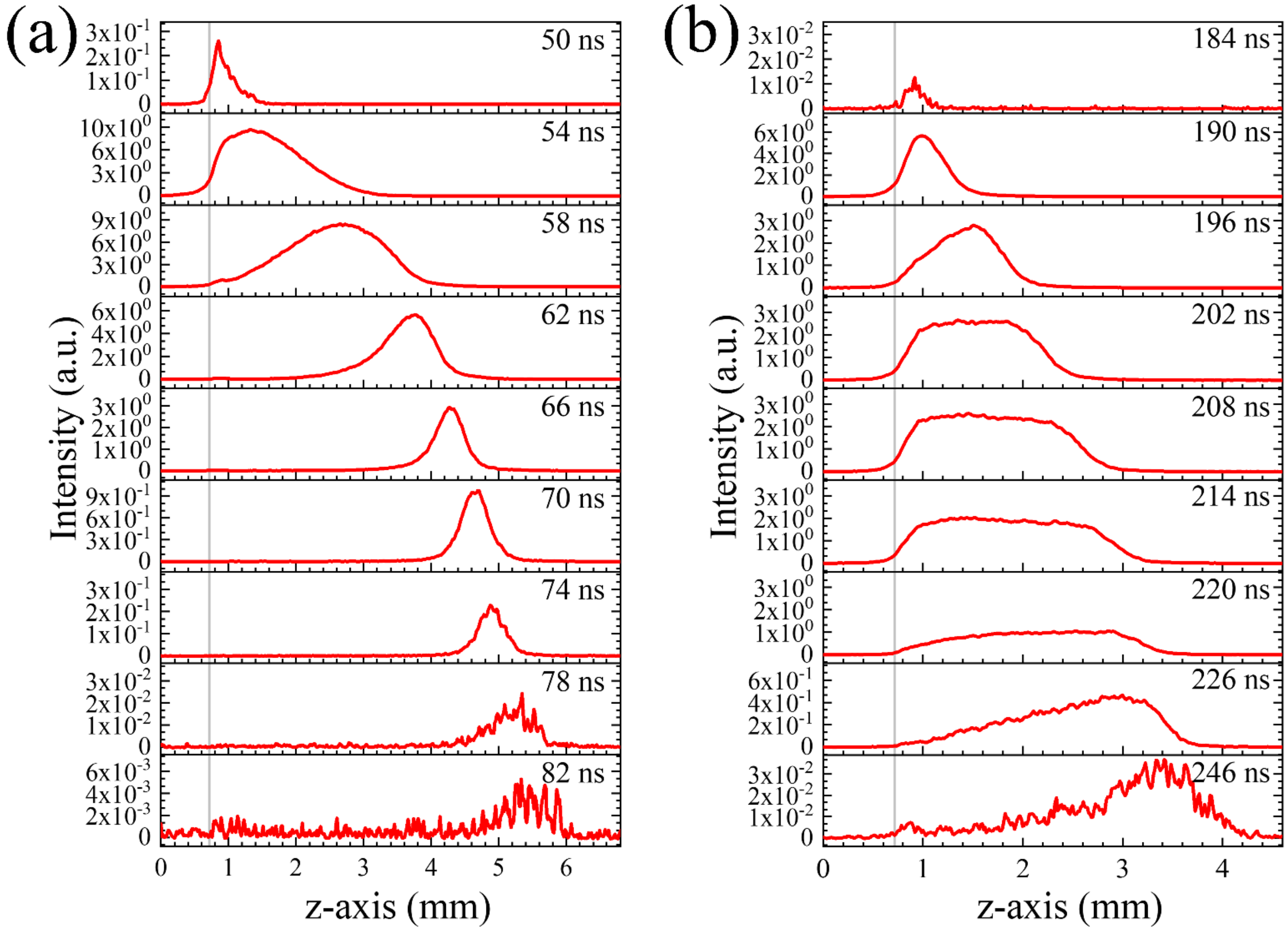

***Figure 10.*** *ns-pulsed operation (9 kV$_p$); see Fig. 4 for time correlation with voltage waveforms. Rows-integrated intensity profiles along z-axis, reduced to the maximal value obtained at 52 ns (not shown here); 2 ns gate, 100 accumulations, 255 gain.* ***(a)*** *Rising part.* ***(b)*** *Falling part. The light gray line denotes the boundary of the electrode gap (section II).*

**Figure 11** visualizes the discharge propagation operating with AC voltage (see **Fig. 4** for time correlation with voltage waveforms). It implies the existence of two simultaneous discharges both during the rising and falling parts of the voltage. On the one hand, as has been shown in **Fig. 9**, a discharge channel prolongates without any obvious separation from the driven electrode. On the other hand, a locally developed discharge coexists, being sustained in the vicinity of the driven electrode (see the intensity peak close to the light gray line in **Fig. 11**). These distinct discharges are formed, intensified and extinguished almost in parallel, while this progress takes about 33 μs during the rising part of the voltage and about 35 μs during the falling part.

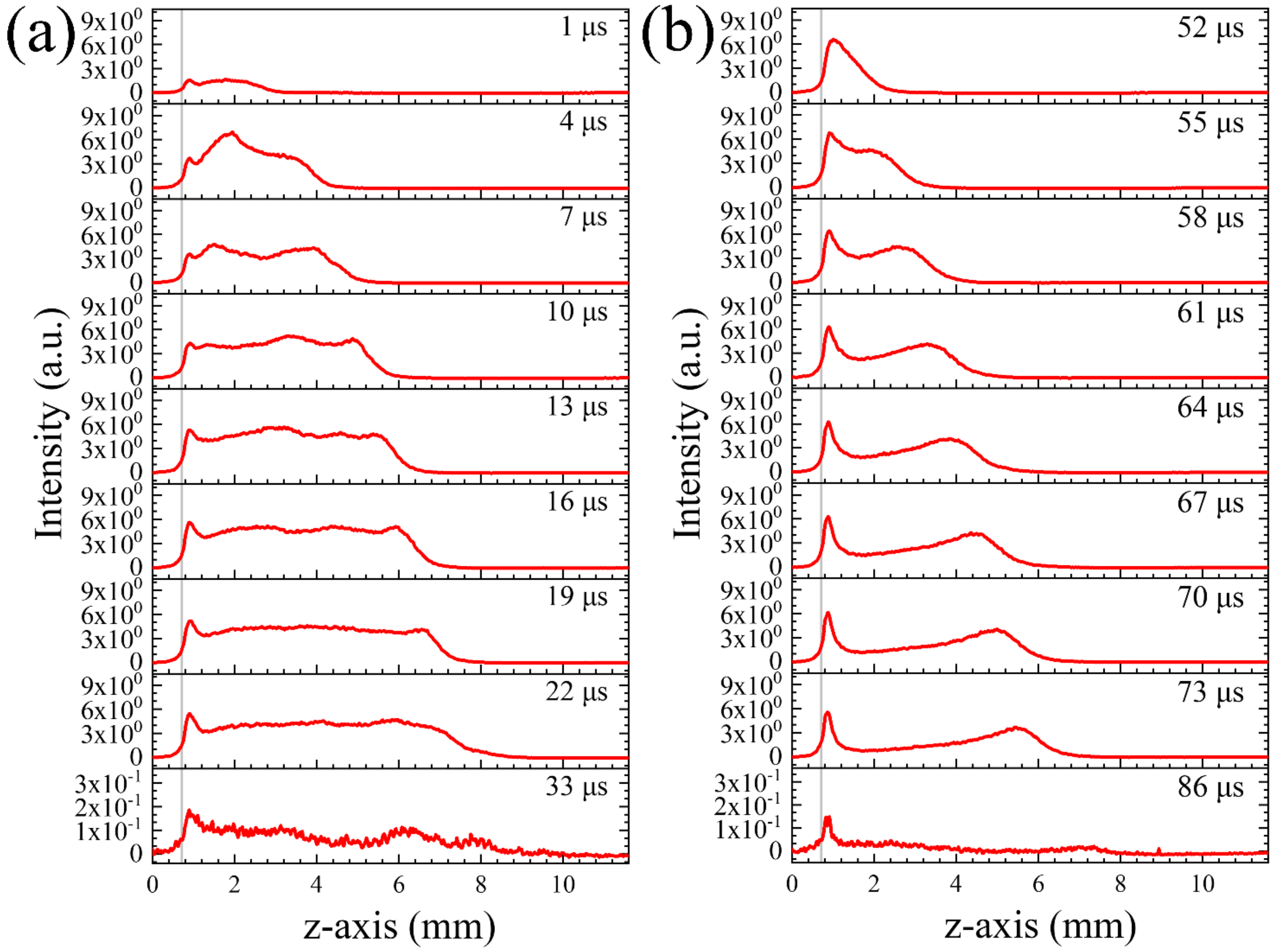

***Figure. 11.*** *AC operation (10 kV$_p$); see Fig. 4 for time correlation with voltage waveforms. Rows-integrated intensity profiles along z-axis, reduced to the maximal value obtained at 2 μs (not shown here); 1 μs gate, 100 accumulations, 255 gain.* ***(a)*** *Rising part.* ***(b)*** *Falling part. The light gray line denotes the boundary of the electrode gap (section II).*

Based on the above results, the precise propagation length and velocity of the SDBD have been assessed as a function of time within both voltage waveforms (**Fig. 12**; see **Fig. 4** for time correlation with voltage waveforms).

In the ns-pulsed operation (**Fig. 12a**), the propagation length increases, mainly at a diminishing rate, both during the rising and falling parts of the voltage, although the increasing rate is constantly higher during the first part. Furthermore, a longer propagation is observed during the rising part compared to the falling part (about 5 mm vs. 3 mm). Correspondingly, the propagation velocity is found to be ~$10^5$ m $s^{-1}$ during both parts of the driving voltage, though the maximal value measured is about 5 times higher during the rising part than in the falling part. These values can be compared with different studies performed on planar SDBDs. As an example, Giotis *et al.* [30] investigated a planar SDBD driven by a pulsed high voltage ($V_p$=10 kV, f=2 kHz, 2 μs pulse width). They measured maximum propagation lengths of about 8 and 3 mm during the rising and falling voltage parts, respectively. The ionization wave velocity during the rising part reached a peak value of about 2×$10^5$ m/s, i.e., about 2 times lower than the corresponding value shown in **Fig. 12a**. The small differences with our results (**Fig. 12a**) can be due to the dissimilar voltage pulse properties and the rather large ICCD gate width (10 ns) used to capture the discharge dynamics in reference [30]. In another study, Benard *et al.* [49] used the same power supply as in this work to drive a planar SDBD. The maximum discharge length during the rising voltage part (4 mm; measured based on ICCD images integrated over the rising part of the voltage) is very close to that obtained on **Fig. 12a**. The corresponding reported maximum velocity is $10^5$ m/s, i.e, about 4 times smaller than in our work. It is not clear if this was measured using an ICCD gate as small as in our case (i.e., 2 ns; **Fig. 12a**). Velocity values during the negative voltage part are not reported in [30,49].

In the AC operation (**Fig. 12b**), once more, the propagation length increases both during the rising and falling parts of the voltage and the increasing rate is higher during the first part. However, the changing rates are not monotonously diminishing, and clear variations are observed. This may be due to physical effects (erratic formation of ionizing channels as discussed above) as well as to the fact that the ICCD gate is here 1 μs wide (i.e., 500 times larger than for the ns-pulsed case) leading rather to "mean" than to "instantaneous" velocities. Thus, the velocities in the AC case appear to be about 3 orders of magnitude smaller than in the ns-pulsed case. Similar velocity measurements under AC operation have been performed in the past using either ICCD cameras [51] or photomultiplier tubes [47,52]. However, the non-averaged velocities of individual cathode-directed surface streamers in the AC case are expected to be of the same order of magnitude as the ns-pulsed SDBDs. This is based on the fact that the temporal properties of the discharge current impulses recorded under ns-pulsed and AC cases are quite similar (**Fig. 4**). These velocities, however, cannot be determined using the lowest available gate width of this ICCD camera (2 ns; as done for the ns-pulsed case) due to their stochastic spatiotemporal development. Grosch *et al.* [53] studied the spatiotemporal dynamics of single AC-driven SDBDs consisting of two needle electrodes placed on a dielectric plate. Using cross-correlation spectroscopy it was possible to perform a direct image of the discharge emission on a spectrometer slit and derive more accurate streamer velocities through the spatiotemporal distribution of the emission intensity of the $N_2^+$(FNS). These measurements provided streamer velocities on the order of a few $10^5$ m/s which are very close to the ones measured for the ns-pulsed case in our work. Therefore, although it is possible to extract qualitative averaged streamer velocities using the ICCD camera in AC-driven SDBDs (**Fig. 12b**), it should be kept in mind that these are not representative of the real streamer velocities.

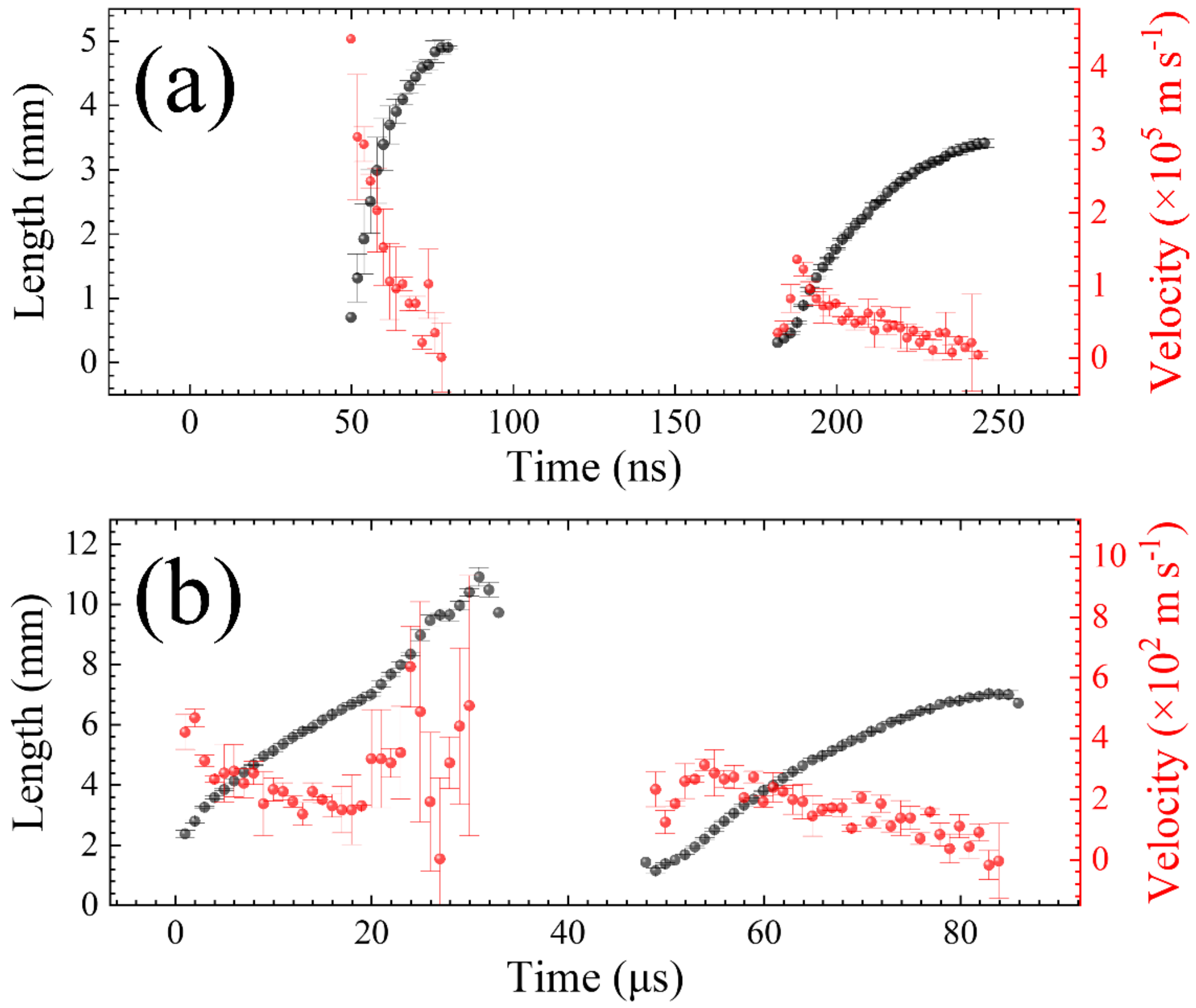

***Figure. 12.*** *SDBD propagation length and velocity; see Fig. 4 for time correlation with voltage waveforms and Figs. 10 and 11 for the ICCD settings.* ***(a)*** *ns-pulsed operation (9 kV$_p$).* ***(b)*** *AC operation (10 kV$_p$).*

Finally, it is important to note that the results presented in **Figs. 10-12** are given with respect to the time points of the voltage–current waveforms plotted in **Fig. 4** (see intervals defined with the dashed lines). This comparison shows that non-zero current values do not necessarily imply the presence of ionization and in turn discharge, since the current consists of a displacement and a drift component (see previous comments on **Fig. 4**). A pronounced case is the comparison between **Fig. 4a** and **Fig. 10a**, where the current begins around 0 ns whereas the discharge emission becomes detectable around 50 ns, respectively.

## V. CONCLUSIONS

The present work proposed the use of quartz barriers, stainless steel electrodes, and transformer oil insulation to implement curved-surface, cold plasma reactors tailored for reliable studies of dielectric barrier discharges propagating onto surfaces. It was demonstrated that such a reactor can efficiently sustain discharges driven either by ns-pulsed or alternative current power sources, in the kHz range, allowing for the determination of mean electrical power consumed ($P$). It was shown that $P$ exhibits a dissimilar non-linear increase with the rising peak voltage ($V_p$) in each case: $P$=5.8×10$^{-4}$×$V_p^{3.8}$ for ns-pulsed (firstly reported in this work) and $P$=2.5×10$^{-5}$×$V_p^{5.4}$ for AC operation. For the conditions studied here, P varies between 0.8 and 2.5 W for the ns-pulsed ($V_p$=7–9 kV) and between 0.9 and 5.3 W ($V_p$=7–10 kV) for the AC operation. Furthermore, discharge dynamics was studied using ICCD imaging. In the ns-pulsed case, the visible pattern of the SDBD consists of well-separated emissive channels starting from zones localized on the driven electrode and propagating onto the dielectric surface downstream of the electrode. This visible pattern is the result of the superposition of distinct ionized channels (resembling cathode-directed streamers) and glow-like discharge (resembling cathode-sheath) both propagating downstream of the driven electrode, whereas the latter remains attached to the electrode. In the AC voltage, the corresponding pattern may be separated into two parts: "cathode-directed streamers spread around the driven electrode and a "glow-like" propagating on the dielectric surface downstream of the electrode. In this case, the pattern is due to erratic, elongated ionization channels in a filamentary form developed during the voltage increase, and to weak ionization occurred during the voltage decrease. Under the present experimental conditions and considerations, the ns-pulsed discharge is associated with ionizing wave propagation appearing to be about three orders of magnitude faster than that of the AC driven discharge. However, the measured velocities in the AC-case were measured using a 500 times larger ICCD gate width. The non-averaged velocities of individual cathode-directed surface streamers extracted in the AC case are expected to be of the same order of magnitude as for the ns-pulsed SDBDs. This is based on the fact that the temporal properties of the discharge current impulses recorded under ns-pulsed and AC cases are quite similar. The present study helps in better understanding ns-pulsed and AC-driven cylindrical SDBDs generated using a prototype sturdy reactor for applications in flow control, biomedicine and agriculture.

**Acknowledgement**

Authors express their acknowledgements to (i) Mr Gerasimos Diamantis (University of Patras) for technical support on the SDBD reactor, (ii) Dr Laurent Invernizzi (LSPM, CNRS) for technical support on the experimental setup, and (iii) Dr Odhissea Gazeli (University of Cyprus) for graphical support on Fig. 6.

This work was funded by the Labex SEAM Project (ANR–10–LABX–0096; ANR–18–IDEX–0001), the ANR ULTRAMAP Project (ANR–22–CE51–0027), the IDF regional Project SESAME DIAGPLAS, and the DFG project d3DBD (535827833).

## ORCID iDs

K. Giotis https://orcid.org/0000-0002-8615-331X

D. Stefas https://orcid.org/0000-0003-2390-4603

Y. Agha https://orcid.org/0009-0006-3932-3781

H. Höft https://orcid.org/0000-0002-9224-4103

X. Duten https://orcid.org/0000-0001-5379-4391

P. Svarnas https://orcid.org/0000-0002-2818-744X

G. Lombardi https://orcid.org/0000-0001-6583-0046

K. Gazeli https://orcid.org/0000-0002-5479-0373

## Data Availability Statement

The data that support the findings of this study are available from the corresponding author upon reasonable request.